\documentclass[aps,prl,reprint,showpacs,superscriptaddress,10pt]{revtex4-2}  
\usepackage{epsfig}
\usepackage{epstopdf}
\usepackage{graphicx}
\usepackage{amsmath,amssymb}
\usepackage{amsmath,bm}
\usepackage{babel}
\usepackage{txfonts}
\usepackage{physics}
\usepackage{comment}
\usepackage{xcolor}
\usepackage{lipsum}
\usepackage[caption=false]{subfig}
\usepackage{lineno,blindtext}
\usepackage{diagbox, eqparbox, hhline}
\usepackage{soul}
\usepackage{xcolor}
\usepackage[colorlinks,citecolor=blue,linkcolor=blue,urlcolor=blue,]{hyperref}
\usepackage{bbm}

\begin{document}

    \title{ 
    Crossover to Sachdev-Ye-Kitaev criticality 
    in an infinite-range quantum  Heisenberg spin glass}

	\author{Hossein Hosseinabadi}
    \email{hossein@pks.mpg.de}
    \altaffiliation{\\Present address: Max Planck Institute for the Physics of Complex Systems, 01187 Dresden, Germany.}
    \affiliation{Institute of Physics, Johannes Gutenberg University Mainz, 55099 Mainz, Germany}

    \author{Subir Sachdev}
    \affiliation{Department of Physics, Harvard University, Cambridge, MA 02138, USA}
    \affiliation{Center for Computational Quantum Physics, Flatiron Institute, New York, NY 10010, USA}

    \author{Jamir Marino}
    \affiliation{Department of Physics, State University of New York at Buffalo, Buffalo, NY 14260, USA}

	\begin{abstract}
		{A central question in frustrated quantum magnets is how quantum fluctuations destabilize spin-glass order and promote quantum spin-liquid behavior.}
        We study the equilibrium dynamics of an infinite-range quantum Heisenberg model with random couplings, in which local magnetic moments arise from $\mathcal{N}_f$ flavors of spinful fermions. We employ an expansion in $\mathcal{N}_f$, which controls the strength of quantum fluctuations, and self-consistently include $1/\mathcal{N}_f$ corrections to the Luttinger–Ward functional. In the large-$\mathcal{N}_f$ limit, where quantum fluctuations are weak, the high- and low-temperature phases are respectively paramagnetic and spin glass ordered, with a transition temperature independent of $\mathcal{N}_f$. For small numbers of fermionic flavors, however, quantum fluctuations substantially suppress the ordering temperature. We show that this behavior reflects the proximity of the system to a Sachdev–Ye–Kitaev (SYK) {regime, realizing quantum spin-liquid dynamics over a broad range of finite frequencies}, where both fermionic and spin spectral densities display critical behavior over a broad range of finite frequencies, with the latter exhibiting the scale-invariant   form  $\chi''(\omega)\sim \operatorname{sgn}(\omega)$. At the lowest energies and temperatures, spin glass dynamics ultimately take over, producing a universal sub-Ohmic dynamical spin susceptibility  $\chi''(\omega)\sim \operatorname{sgn}(\omega)\sqrt{|\omega|}$. {Our results highlight a direct connection between the suppression of spin-glass order and the emergence of SYK criticality in a frustrated quantum magnet.}
	\end{abstract}
	
	\maketitle

    The interplay of frustration and fluctuations gives rise to a variety of phases in interacting quantum many-body systems~\cite{Sachdev_2011,Sachdev_2023}. Among the most prominent examples are spin glasses (SG) and quantum spin liquids (QSL). SG are ordered phases in which magnetic moments freeze into a disordered configuration at low temperatures; consequently, both thermal and quantum fluctuations tend to destabilize them~\cite{Binder_Young_1986,mezard1988spin}. In contrast, QSL remain paramagnetic (PM) without a local order parameter, but exhibit strong entanglement and unconventional low-energy excitations with distinct properties such as fractionalization~\cite{Savary_Balents_2017,Zhou_Kanoda_Ng_2017,Knolle_Moessner_2019}.

    A potential theoretical route toward realizing a QSL is to start from a SG and enhance fluctuations so as to suppress the freezing of spins. In this picture, sufficiently strong thermal or quantum fluctuations can melt SG order and give rise to a strongly fluctuating PM state. Evidence for such behavior has emerged in studies of long-range quantum Heisenberg spin glasses and related models, including extensions to SU($M$) spins~\cite{Sachdev_Ye_1993,Georges_Parcollet_Sachdev_2000,Georges_Parcollet_Sachdev_2001,Christos_Haehl_Sachdev_2022,Chowdhury_Georges_Parcollet_Sachdev_2022}, and systems with mobile spins~\cite{Joshi_Li_Tarnopolsky_Georges_Sachdev_2020,Shackleton_Wietek_Georges_Sachdev_2021}, which point to the appearance of QSL regimes either as ground states or over finite ranges of temperature or frequency. 

    \begin{figure}[!t]
        \centering
        \hspace{-15pt}\includegraphics[width=0.85
        \linewidth]{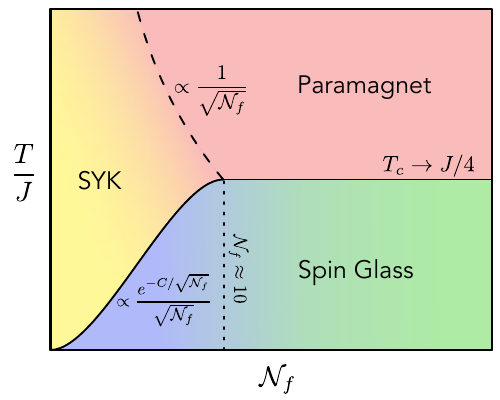}
        \caption{{Schematic phase diagram versus $T$ and $\mathcal{N}_f$. At large $\mathcal{N}_f$, $T_c\to J/4$. $\mathcal{N}_f\approx10$ marks the approximate onset of important SYK fluctuations. Small-$\mathcal{N}_f$ asymptotic scalings are analytical, while the curves are schematic.}}
        \label{fig:phase_diag}
    \end{figure}

    {In this work we introduce a fermionic formulation in which the strength of quantum fluctuations can be tuned directly through the flavor number, allowing us to track the evolution from spin-glass ordering to an SYK-dominated regime within a single model.} We study a fully connected quantum Heisenberg model in which the spins are represented by $\mathcal{N}_f$ flavors of fermions, with the strength of quantum fluctuations   tuned by the flavor number. Using a Luttinger-Ward functional approach~\cite{Rammer_2007}, we evaluate the behavior of the model across temperature and $\mathcal{N}_f$. {Accordingly, at large $\mathcal{N}_f$, where quantum fluctuations are suppressed, the system exhibits a trivial PM phase at high temperatures and a spin-glass transition whose temperature saturates with increasing $\mathcal{N}_f$, as shown in Fig.~\ref{fig:phase_diag}.} In contrast, at small $\mathcal{N}_f$ quantum fluctuations strongly weaken the SG order, leading to an exponential suppression of the transition temperature with decreasing $\mathcal{N}_f$. {We attribute this behavior to the emergence of a strongly fluctuating QSL regime, manifested through Sachdev–Ye–Kitaev (SYK) critical behavior over a finite range of frequencies~\cite{Chowdhury_Georges_Parcollet_Sachdev_2022}}, which is also reflected in the structure of the low-energy fermionic and magnetic spectra. Notably, the latter exhibits a plateau over a wide range of frequencies, suggesting the emergence of a finite-frequency QSL phase~\cite{Sachdev_Ye_1993,Chowdhury_Georges_Parcollet_Sachdev_2022,Christos_Haehl_Sachdev_2022}. At lower temperatures, critical fermions eventually freeze into a SG phase, where we observe a crossover of the spin spectral density to sub-Ohmic behavior at small frequencies.
    ~\\

	\emph{Model.---}We consider a fully connected network of fermions with random exchange interactions.
    The fundamental degrees of freedom are $\mathcal{N}_f$ flavors of spin-$1/2$ fermions
    $\psi_{i\lambda s}$ on each of the $N$ sites, with $\lambda=1,\dots,\mathcal{N}_f$
    and $s=\uparrow,\downarrow$. The Hamiltonian describes random SU(2)-invariant exchange between
    collective spin densities,
    \begin{equation}
    H = \frac{1}{\mathcal{N}_f\sqrt{N}} \sum_{i<j} J_{ij} \, \mathbf{S}_i \cdot \mathbf{S}_j ,
    \label{eq:H}
    \end{equation}
    where the exchange couplings $J_{ij}$ are independent Gaussian variables
    with $\overline{J_{ij}}=0$ and $\overline{J_{ij}^2}=J^2$.
    The $1/\sqrt{N}$ scaling ensures a well-defined thermodynamic limit,
    while the prefactor $1/\mathcal{N}_f$ renders the interaction energy extensive in
    the number of fermionic flavors, allowing for a controlled large-$\mathcal{N}_f$ expansion~\cite{SupplementalMaterial}.  The local spin operators are collective fermion bilinears,
    \begin{equation}
    \mathbf{S}_i =
    \frac{1}{2} \sum_{\lambda=1}^{\mathcal{N}_f}
    \sum_{s,s'=\uparrow,\downarrow}
    \psi^\dagger_{i\lambda s} \,
    \boldsymbol{\sigma}_{ss'} \,
    \psi_{i\lambda s'} ,
    \label{eq:Si}
    \end{equation}
    which obey the SU(2) algebra. {Defining the normalized collective spin $\mathbf{m}_i=\mathbf{S}_i/\mathcal{N}_f$, its commutators scale as $[m_i^\alpha,m_i^\beta]=i\epsilon_{\alpha\beta\gamma}m_i^\gamma/\mathcal{N}_f$, such that the large-$\mathcal{N}_f$ limit becomes increasingly semiclassical.} Throughout this work, we focus on the half-filled regime, 
    $\langle \psi^\dagger_{i\lambda s}\psi_{i\lambda s}\rangle = 1/2$, 
    which minimizes the density of empty or doubly-occupied sites and thereby enhances magnetic fluctuations.

    The model is structurally similar to the Heisenberg SG~\cite{Bray_Moore_1980}. However, in contrast to conventional Heisenberg models where spins are restricted to a fixed SU(2) representation, here the magnitude of spins is not fixed, and the local moment $\langle \mathbf{S}_i^2 \rangle$ can fluctuate. As a result, the dependence of phase boundaries on $\mathcal{N}_f$ differs from that of the Heisenberg SG with fixed-length spins~\cite{Bray_Moore_1980,Sachdev_Ye_1993,Georges_Parcollet_Sachdev_2000,Georges_Parcollet_Sachdev_2001}. As we show later, the present model also share some of its features with metallic SG~\cite{Sachdev_Read_Oppermann_1995,Sengupta_Georges_1995,Pastor_Dobrosavljević_1999,Müller_Strack_Sachdev_2012}.

    The quantities of interest are the disorder-averaged fermionic and spin response functions, defined respectively as
    \begin{align}
        G^R(t)\delta_{ij}\delta_{\lambda \lambda'}\delta_{ss'}&=-i\Theta(t) \Big\langle \big[ \psi_{i\lambda s}(t),\psi^\dagger_{j\lambda s'}(0)\big]_+\Big\rangle,\label{eq:GR_def}\\
        \chi(t)\delta_{ij}\delta_{\alpha \beta} &= -\frac{i}{\mathcal{N}_f}\Theta(t) \Big\langle \big[ S_i^\alpha(t),S_j^\beta(0)\big]_-\Big\rangle,\label{eq:chi_def}
    \end{align}
    as well as the symmetric spin correlation function
    \begin{equation}\label{eq:C_def}
        C(t)\delta_{ij}\delta_{\alpha \beta}= \frac{1}{\mathcal{N}_f}\Big\langle \big[ S^\alpha_i(t),S^\beta_j(0)\big]_+ \Big\rangle,
    \end{equation}
    where $[A,B]_\pm = AB \pm BA$. Throughout, expectation values are taken with respect to both the quantum state and the disorder ensemble. We have also used the fact that, in the thermodynamic limit, these correlators become diagonal in all indices due to the infinite-range nature of the interactions~\cite{Bray_Moore_1980}.
    \\~
    
    \emph{Method.---}We employ the Keldysh formalism, which provides direct access to the two-point functions defined in Eqs.~\eqref{eq:GR_def}–\eqref{eq:C_def} and facilitates the averaging over quenched disorder~\cite{Rammer_2007,kamenev2023field,Hosseinabadi_short2024,Hosseinabadi_long2024}. As detailed in the Supplemental Material~\cite{SupplementalMaterial}, we derive the Luttinger-Ward (LW) functional $\Gamma[G]$ for the system to next-to-leading order in an expansion in powers of $\mathcal{N}_f$~\cite{Babadi_Demler_Knap_2015,Hosseinabadi_short2024,Hosseinabadi_long2024,Mikheev_Hosseinabadi_Marino_2025,SupplementalMaterial}. Even at this finite order, the LW functional contains an infinite number of contributions, which can be re-summed via a Hubbard-Stratonovich transformation in terms of collective spin fields.

    \begin{figure}[!t]
        \centering
    \includegraphics[width=0.9\linewidth]{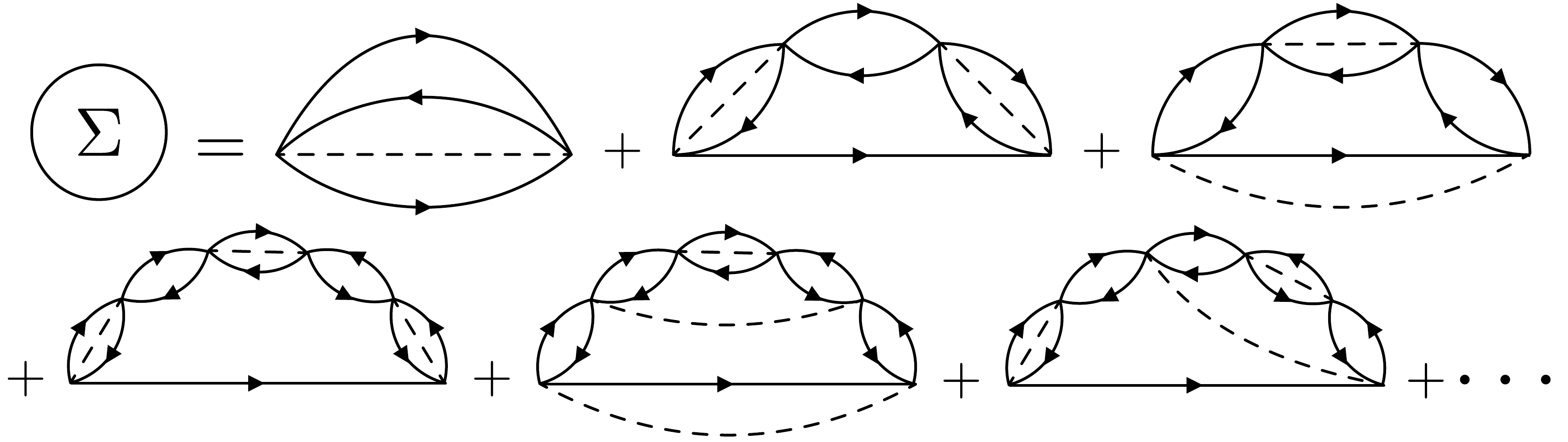}
        \caption{Diagrammatic contributions to the fermionic self-energy,  at leading non-vanishing order in the $1/\mathcal{N}_f$ expansion, include an infinite series which is resummed in this work.}
        \label{fig:self-energy}
    \end{figure}

    \begin{figure*}[!t]
        \centering
        \includegraphics[height=0.24\textwidth]{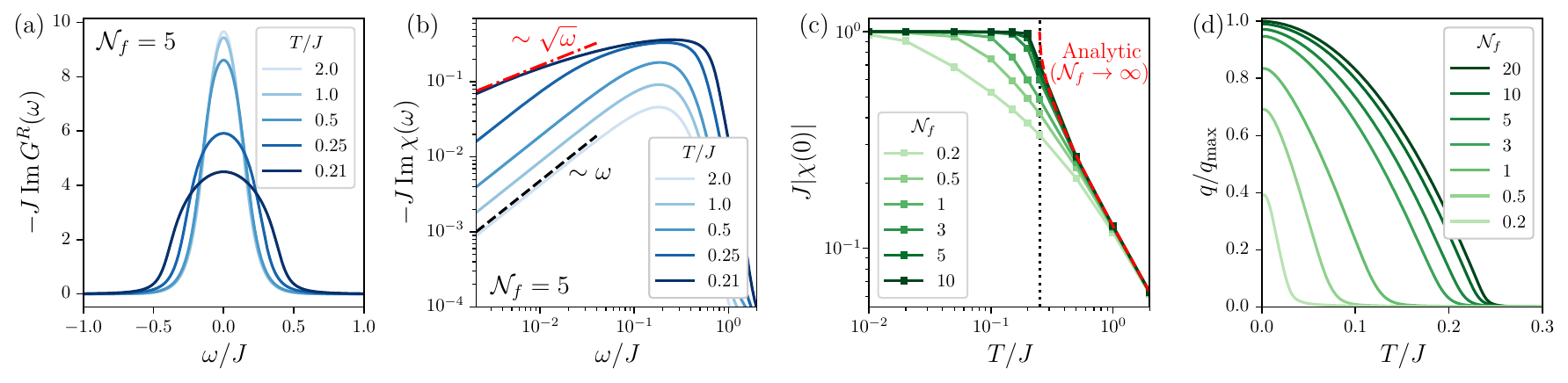}
        \caption{Numerical solutions of the self-consistent equations. (a) Fermionic spectral density for $\mathcal{N}_f=5$ in the SG ($T/J=0.21$) and PM (rest of the curves) phases. (b) Spin spectral density for $\mathcal{N}_f=5$ limit, showing Ohmic (black dashed) and sub-Ohmic (red dash-dotted) low-energy behavior in the PM and SG phases, respectively. (c) Temperature dependence of the static magnetic susceptibility for different values of $\mathcal{N}_f$. The red curve shows the analytical result obtained for $\mathcal{N}_f\to\infty$. (d) SG order parameter as a function of temperature for different values of $\mathcal{N}_f$, {normalized by $q_{\max}$, the numerical value of $q$ at $T=0$ and $\mathcal{N}_f\to\infty$.}}
        \label{fig:panel}
    \end{figure*}
    
    We work in the frequency domain, where the fermionic response function takes the form $G^R(\omega)=(\omega-\Sigma^R(\omega))^{-1}$, with $\Sigma^R(\omega)$ the self-energy. Figure~\ref{fig:self-energy} shows the diagrammatic expansion of the self-energy within our approximation scheme. Upon evaluating the diagrams, one obtains
\begin{equation}\label{eq:SigR_def}
    \Sigma^R(\omega)=\frac{3i}{8\mathcal{N}_f}\int_{-\infty}^{+\infty} \Big(W^K(\nu)G^R(\omega-\nu)+ W^R(\nu)G^K(\omega-\nu)\Big)\frac{d\nu}{2\pi},
\end{equation}
where $G^K(\omega)$ is the fermionic Keldysh Green’s function. It is related to $G^R(\omega)$ via the fluctuation-dissipation theorem according to $G^K(\omega)=2i\Im G^R(\omega)\tanh(\omega/2T)$~\cite{kamenev2023field}. In Eq.~\eqref{eq:SigR_def}, $W^{R/K}(\omega)$ denote the retarded and Keldysh components of the correlation function of the field experienced by the spins, given by
\begin{align}
    W^R(\omega)&=\frac{J^2 \chi(\omega)}{1-J^2\Pi^R(\omega)\chi(\omega)},\\
    W^K(\omega)&=\frac{-iJ^2 C(\omega)}{\abs{1-J^2\Pi^R(\omega)\chi(\omega)}^2}+J^4\abs{\chi(\omega)}^2\Pi^K(\omega).
\end{align}
Here, $\Pi^R(\omega)$ is the polarization function,
\begin{equation}
    \Pi^R(\omega)=-\frac{i}{4}\int_{-\infty}^{+\infty} \Big(G^K(\nu)G^A(\nu-\omega)+ G^R(\nu)G^K(\nu-\omega)\Big)\frac{d\nu}{2\pi},
\end{equation}
where $G^A=(G^R)^*$, and $\Pi^K(\omega)=2i\Im \Pi^R(\omega)\coth(\omega/2T)$. This set of equations is closed by the expressions for the spin response and correlation functions,
\begin{equation}\label{eq:chi_w}
    \chi(\omega)=\frac{\Pi^R(\omega)}{1-J^2 \Pi^R(\omega)\chi(\omega)},
\end{equation}
and
\begin{equation}\label{eq:C_w}
    \Big(1 - J^2|\chi(\omega)|^2 \Big)\,C(\omega) = \frac{i\Pi^K(\omega)}{|1-J^2\Pi^R(\omega)\chi(\omega)|^2}.
\end{equation}
The structure of the diagrams in Fig.~\ref{fig:self-energy} closely resembles that of the GW approximation~\cite{Aryasetiawan_1998}, with the key difference that disorder averaging permits multiple ways of drawing disorder lines, changing the combinatorial prefactors.

    {In the following, we analyze the self-consistent equations analytically at low energies and numerically over the full frequency domain. The resulting behavior is summarized in Fig.~\ref{fig:phase_diag}.}
    %In the following, we analyze the self-consistent solution of the above equations in different parameter regimes, using a combination of analytical arguments at low energies and numerically exact solutions over the full frequency domain. The resulting behavior is summarized in the phase diagram shown in Fig.~\ref{fig:phase_diag}. We start from the PM phase as the simplest phase of the model, and then discuss its transition into a SG. This is followed by the discussion of the SYK phase and its corresponding SG transition.
    \\~

    \emph{Paramagnet.---}We begin with the thermally dominated PM phase, corresponding to the regime in which the temperature $T$ exceeds all local energy scales, including the coupling $J$ and the fermionic spectral broadening. In this limit, the static magnetic susceptibility can be obtained from elementary statistical-mechanics considerations, yielding $\chi(0)=-1/8T$, consistent with Curie’s law~\cite{blundell2001magnetism}. {This observation suggests that $J\chi(\omega)\sim J/T$ can serve as a small parameter for a perturbative expansion in the high-temperature regime.} To leading order, this expansion yields $\chi(\omega)\approx \Pi^R(\omega)$ and $W^R(\omega)\approx J^2\Pi^R(\omega)$, corresponding to retaining only the first term in the self-energy expansion shown in Fig.~\ref{fig:self-energy}.

    Even with this simplification, a full analytical solution of the equations in the entire frequency domain remains inaccessible. {At low-frequencies an ansatz of the form $G^R(\omega)\approx(\omega - i\gamma)^{-1}$ correctly reproduces the static susceptibility and self-consistently yields~\cite{SupplementalMaterial}}
    \begin{equation}\label{eq:spec_PM}
        \gamma  =\frac{\sqrt{3}J}{8\mathcal{N}_f^{1/2}}, \quad \Im \chi(\omega)\approx -\frac{\omega}{16\gamma T}.
    \end{equation}
    {The corresponding numerical fermionic and spin spectral densities are shown in Figs.~\ref{fig:panel}(a)~and~\ref{fig:panel}(b), respectively.} We see that at high temperatures, the fermionic broadening $\gamma$ is insensitive to $T$ and is instead set by an emergent energy scale proportional to $J\mathcal{N}_f^{-1/2}$. The spin excitations exhibit an Ohmic low-energy spectrum, whose magnitude increases with $\mathcal{N}_f$ due to the enlarged phase space for spin-exchange processes as the number of fermion flavors grows. Up to numerical prefactors, Eq.~\eqref{eq:spec_PM} is in good agreement with the numerical results in Figs.~\ref{fig:panel}(a) and~\ref{fig:panel}(b) at high temperatures.
    \\~

    \emph{Spin glass.---}Upon lowering the temperature from the PM phase, the spin susceptibility becomes enhanced. As a result, $J|\chi(0)|$ is no longer small, and the system becomes unstable toward the formation of SG order. The latter is characterized by finite spin correlations (Eq.~\eqref{eq:C_def}) at long times, such that $\lim_{t\to\infty} C(t)>0$~\cite{Sompolinsky_Zippelius_1981,Binder_Young_1986}. In the frequency domain, this behavior corresponds to a singular contribution of the form
    \begin{equation}\label{eq:q_def}
        C(\omega)=4\pi \mathcal{N}_f q \delta(\omega)+\tilde{C}(\omega),
    \end{equation}
    where $q$ is the Edwards-Anderson order parameter~\cite{Sompolinsky_Zippelius_1981,Binder_Young_1986}, which is finite in the SG phase, and $\tilde{C}(\omega)$ denotes the regular part of $C(\omega)$. $\tilde{C}(\omega)=-2\Im \chi(\omega) \coth(\omega/2T)$. {Here $q$ is normalized to remove the trivial $\mathcal{N}_f$ dependence of the collective spin magnitude.} Substituting Eq.~\eqref{eq:q_def} into Eq.~\eqref{eq:C_w} and integrating over an infinitesimal frequency window around $\omega=0$, we obtain $q(1-J^2\chi^2(0))=0$. For $q>0$, this implies
    \begin{equation}\label{eq:Stoner}
       J|\chi(0)|=1,
    \end{equation}
    which is formally analogous to the Stoner criterion~\cite{blundell2001magnetism}. In contrast to the conventional case, however, $|\chi(0)|$ remains pinned to $J^{-1}$ throughout the entire SG phase, a condition known as marginal stability~\cite{Miller_Huse_1993,Sachdev_Read_Oppermann_1995}.

    To determine the critical temperature, we use the fact that $\Pi^R(0)$ is still given by $\Pi^R(0)=-1/8T$ as long as $\gamma \lesssim T$, which is valid for sufficiently large $\mathcal{N}_f$. Substituting this expression into Eq.~\eqref{eq:chi_w}, we obtain
    \begin{equation}\label{eq:chi0_pm}
        \chi(0)/4T= \sqrt{1-(J/4T)^2}-1 \quad \to \quad T_c= J/4,
    \end{equation}
    in good agreement with the numerical solution shown in Fig.~\ref{fig:panel}(c). The value of the order parameter is determined by imposing Eq.~\eqref{eq:Stoner}, whose numerical solution is shown in Fig.~\ref{fig:panel}(d), demonstrating that SG order emerges continuously below $T_c$. SG order gives rise to an effective contribution to the fermionic self-energy~\cite{SupplementalMaterial},
    \begin{equation}\label{eq:Sig_SG}
        \Sigma^R_\mathrm{SG}(\omega)\approx 3J^2 q\,G^R(\omega),
    \end{equation}
    which is formally analogous to the self-energy of fermions subject to static disorder of effective strength $\mathcal{W}=(3q)^{1/2}J$~\cite{Altland_Simons_2023,Chowdhury_Georges_Parcollet_Sachdev_2022}, reflecting the presence of a randomly frozen background in the SG phase. 
    
    The spin spectral density in the ordered phase can be obtained from Eqs.~\eqref{eq:chi_w} and~\eqref{eq:Stoner}, yielding $\Im \chi(\omega)\sim J^{-1/2}|\Im \Pi^R(\omega)|^{1/2}$ at low frequencies. A general argument, relying only on the regularity of the fermionic spectral density~\cite{SupplementalMaterial}, implies that $\Pi^R(\omega)\approx -\alpha \omega$ at low frequencies throughout the SG phase. This leads to a sub-Ohmic spin spectral density of the form
    \begin{equation}
        \Im \chi(\omega)=-\Big(\frac{2\alpha}{J}\Big)^{1/2} |\omega|^{1/2}\mathrm{sgn}(\omega), \quad T\le T_c,
    \end{equation}
    in agreement with the numerical results shown in Fig.~\ref{fig:panel}(b). Despite the universal sub-Ohmic profile, the coefficient $\alpha$ behaves differently near the transition and deep in the ordered phase. Close to the transition, $\Pi^R(\omega)$ is dominated by thermal fluctuations, similar to Eq.~\eqref{eq:spec_PM}. In contrast, deep in the ordered phase, $\Pi^R(\omega)$ arises from particle-hole excitations near the Fermi surface, {and $\alpha \approx 1/4\pi\mathcal{W}^2$ is controlled solely by the fermionic density of states at the Fermi level}~\cite{SupplementalMaterial}.

    In the discussion above, we focused on the large-$\mathcal{N}_f$ limit, where the critical temperature $T_c$ is independent of $\mathcal{N}_f$. For smaller values of $\mathcal{N}_f$, SG order is progressively weakened due to enhanced quantum fluctuations, as evidenced by the reduced transition temperature and the smaller values of $q$ shown in Figs.~\ref{fig:panel}(c) and~\ref{fig:panel}(d). In the following, we show that this suppression can be understood as a consequence of the proximity of the system to an SYK phase at small $\mathcal{N}_f$. 
    \\~

    \emph{SYK paramagnet.---}Upon decreasing $\mathcal{N}_f$ in the PM phase, we reach a regime in which the fermionic spectral broadening (denoted by $\gamma$ in Eq.~\eqref{eq:spec_PM}) is no longer much smaller than $T$, and thermal fluctuations no longer dominate. Provided that magnetic fluctuations associated with SG ordering remain weak, such that $J|\chi(0)|$ is small, higher-order diagrams in Fig.~\ref{fig:self-energy} can still be neglected. {As shown explicitly in the Supplemental Material~\cite{SupplementalMaterial}, the resulting self-energy reduces in the time domain to that of the complex SYK model $\Sigma(t,t')\approx J_\mathrm{SYK}^2\big[G(t,t')\big]^2G(t',t)$~\cite{Gu_ComplexSYK2020,Chowdhury_Georges_Parcollet_Sachdev_2022}, where $G(t,t')$ denotes the fermionic Green's function on the Keldysh time-contour} and
    \begin{equation}
        J_\mathrm{SYK}= \frac{\sqrt{3}}{4}\frac{J}{\sqrt{\mathcal{N}_f}},
    \end{equation}
    is the effective SYK coupling.
    
\begin{figure}[!t]
        \centering
        \includegraphics[height=.24\textwidth]{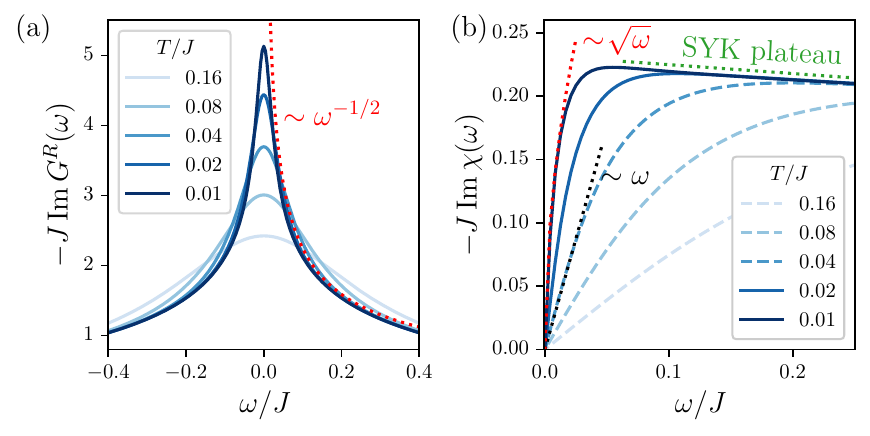}
        \caption{{The emergence of SYK physics. (a) Fermion spectrum developing SYK power-law behavior. (b) Spin spectrum crossing from an intermediate-frequency SYK plateau to low-frequency Ohmic PM (dashed) and sub-Ohmic SG (solid) behavior. In both plots $\mathcal{N}_f=1/10$.}}
        \label{fig:panel_syk}
    \end{figure}

    One may therefore anticipate the emergence of SYK physics for $T\lesssim J_\mathrm{SYK}$. However, the system should remain in the PM phase such that $J|\chi(0)|$ is small. This requires the existence of a finite temperature window satisfying $T_c \lesssim T \lesssim J_\mathrm{SYK}$. Using the large-$\mathcal{N}_f$ expression for $T_c$ given by Eq.~\eqref{eq:chi0_pm}, we obtain an approximate upper bound on $\mathcal{N}_f$,
    \begin{equation}\label{eq:SYK_bound}
        T_c\lesssim J_\mathrm{SYK} \quad \to \quad \mathcal{N}_f\lesssim 10.
    \end{equation}
    Consistent with this estimate, our numerical results in Figs.~\ref{fig:panel}(c) and~\ref{fig:panel}(d) show a suppression of the SG transition in the same range of $\mathcal{N}_f$. The presence of the SYK phase becomes manifest upon examining the fermionic spectral density at small $\mathcal{N}_f$, shown in Fig.~\ref{fig:panel_syk}(a), where $\Im G^R(\omega)\sim |\omega|^{-1/2}$ over a finite frequency window. {We note that although the microscopic model is defined for integer $\mathcal{N}_f$, we formally analytically continue the self-consistent equations to non-integer values of $\mathcal{N}_f$ to make the emergence of SYK criticality particularly transparent, thereby demonstrating the SYK regime underlying the suppression of $T_c$ already observed for $1\lesssim\mathcal N_f\lesssim10$.}
    \\~

    \emph{SYK to spin glass transition.---}The SYK phase ultimately undergoes a transition to a SG state at low temperatures. However, the onset of ordering is strongly suppressed by quantum fluctuations. To demonstrate this, we invoke the Stoner criterion (Eq.~\eqref{eq:Stoner}). In this regime, the static polarization $\Pi^R(0)$ acquires a logarithmic temperature dependence as $\Pi^R(0)\sim J_\mathrm{SYK}^{-1}\ln\left(T/J_\mathrm{SYK}\right)$~\cite{Gu_ComplexSYK2020,Chowdhury_Georges_Parcollet_Sachdev_2022}. Combining this result with Eqs.~\eqref{eq:chi_w} and~\eqref{eq:Stoner}, we obtain the following scaling for the critical temperature,
    \begin{equation}\label{eq:Tc_SYK}
        T_c\sim \frac{J}{\sqrt{\mathcal{N}_f}}e^{-C/\sqrt{\mathcal{N}_f}},
    \end{equation}
    where $C$ is a numerical constant. This result demonstrates {the exponential suppression of SG ordering temperature in the SYK phase, in agreement with Figs.~\ref{fig:panel}(c) and~\ref{fig:panel}(d).} This behavior closely parallels the exponential dependence of the superconducting transition temperature on the fermionic density of states in BCS theory~\cite{Altland_Simons_2023}, where $T_c\sim \exp(-1/U\rho_0)$, with $U$ the pairing interaction and $\rho_0$ the density of states at the Fermi surface. {A conceptually similar suppression of the superconducting transition temperature from an SYK-critical parent state was reported in Ref.~\cite{Esterlis_2019}.}

    The SG phases at large and small values of $\mathcal{N}_f$ also differ qualitatively in the structure of their magnetic excitations. As shown in Fig.~\ref{fig:panel_syk}(b), in the SYK-SG phase the spin spectral density exhibits a plateau extending down to low frequencies~\cite{Chowdhury_Georges_Parcollet_Sachdev_2022}. Below a characteristic frequency scale $\omega^{\star}$, we observe a crossover to sub-Ohmic behavior, $\Im \chi(\omega)\sim |\omega|^{1/2}$, similar to that found in the large-$\mathcal{N}_f$ limit. The crossover scale can be estimated in the zero-temperature limit by comparing the fermionic broadening induced by the SG order (Eq.~\eqref{eq:Sig_SG}) with the SYK self-energy, $\Sigma^R_\mathrm{SYK}(\omega)\sim (J_\mathrm{SYK}|\omega|)^{1/2}$~\cite{Chowdhury_Georges_Parcollet_Sachdev_2022}, yielding
    \begin{equation}
        \Sigma^R_\mathrm{SG}(\omega^{\star})\sim \Sigma^R_\mathrm{SYK}(\omega^{\star}) \quad \to \quad \omega^{\star}\sim J\mathcal{N}_f^{1/2}\,q.
    \end{equation}
    The dependence on $q$ qualitatively agrees with the results of Ref.~\cite{Christos_Haehl_Sachdev_2022} for the Heisenberg model with SU($M$) spins. The crossover from SYK at intermediate frequencies to SG at low frequencies has also been suggested by a numerically-exact solution in Ref.~\cite{Shackleton_Wietek_Georges_Sachdev_2021} for a random $t$-$J$ model. {Our results are consistent with Ref.~\cite{Kavokine_2024}, where a sub-Ohmic spectrum persists down to very low temperatures, particularly in the doped metallic spin-glass regime, while the half-filled Heisenberg spin glass shows a tendency toward Ohmic behavior as $T\to0$. In this respect, our model may be more closely analogous to the doped case, since the local spin magnitude is not fixed and can fluctuate.}
    \\~

\emph{Perspective.---} While we discussed various physical characteristics of the model in both paramagnetic and spin glass phases, an interesting direction for future work concerns the replica structure of the theory in the spin glass phase using the imaginary-time formalism~\cite{mezard1988spin,Read_Sachdev_Ye_1995,Lang_2024}. Moving beyond equilibrium, the conserving Keldysh–LW framework employed here provides a natural route to investigate far-from-equilibrium dynamics of the model~\cite{Cugliandolo_Lozano_1999,Biroli_Parcollet_2002,Hosseinabadi_short2024,Hosseinabadi_long2024,bera2025sachdev,rodriguez2022far}. In particular, it would allow the study of aging phenomena in the glass phase, as well as the relaxation dynamics of both fermionic and spin degrees of freedom. The interplay between slow glassy and fast-thermalizing dynamics of {the SYK quantum spin-liquid regime} gives rise to regimes characterized by distinct short- and long-time behavior, providing a rich setting for exploring non-equilibrium dynamics in strongly interacting quantum systems, both isolated or coupled to an external phononic bath~\cite{Eberlein_Kasper_Sachdev_Steinberg_2017,Haldar_Haldar_Bera_Mandal_Banerjee_2020,Chowdhury_Georges_Parcollet_Sachdev_2022,Hosseinabadi_Kelly_Schmalian_Marino_2023,kelly2021effect}.
~\\

\emph{Acknowledgments---} SS and JM acknowledge the hospitality of ICTP-SAIFR (São Paulo), where this work was initiated. SS was supported by NSF Grant DMR-2245246 and by the Simons Collaboration on Ultra-Quantum Matter which is a grant from the Simons Foundation (651440).  JM acknowledges support from the CAS Dean's office at SUNY Buffalo.

\bibliography{Refs}

\end{document}

\title{Supplemental Material:\\Crossover to Sachdev-Ye-Kitaev criticality 
    in an infinite-range quantum Heisenberg spin glass}

	\author{Hossein Hosseinabadi}
    \email{hossein@pks.mpg.de}
    \affiliation{Institute of Physics, Johannes Gutenberg University Mainz, 55099 Mainz, Germany}

    \author{Subir Sachdev}
    \affiliation{Department of Physics, Harvard University, Cambridge, MA 02138, USA}
    \affiliation{Center for Computational Quantum Physics, Flatiron Institute, New York, NY 10010, USA}

    \author{Jamir Marino}
    \affiliation{Department of Physics, State University of New York at Buffalo, Buffalo, NY 14260, USA}

	\maketitle

    \section{Keldysh Path integral}

\begin{figure}
        \centering
        \includegraphics[height=0.14\linewidth]{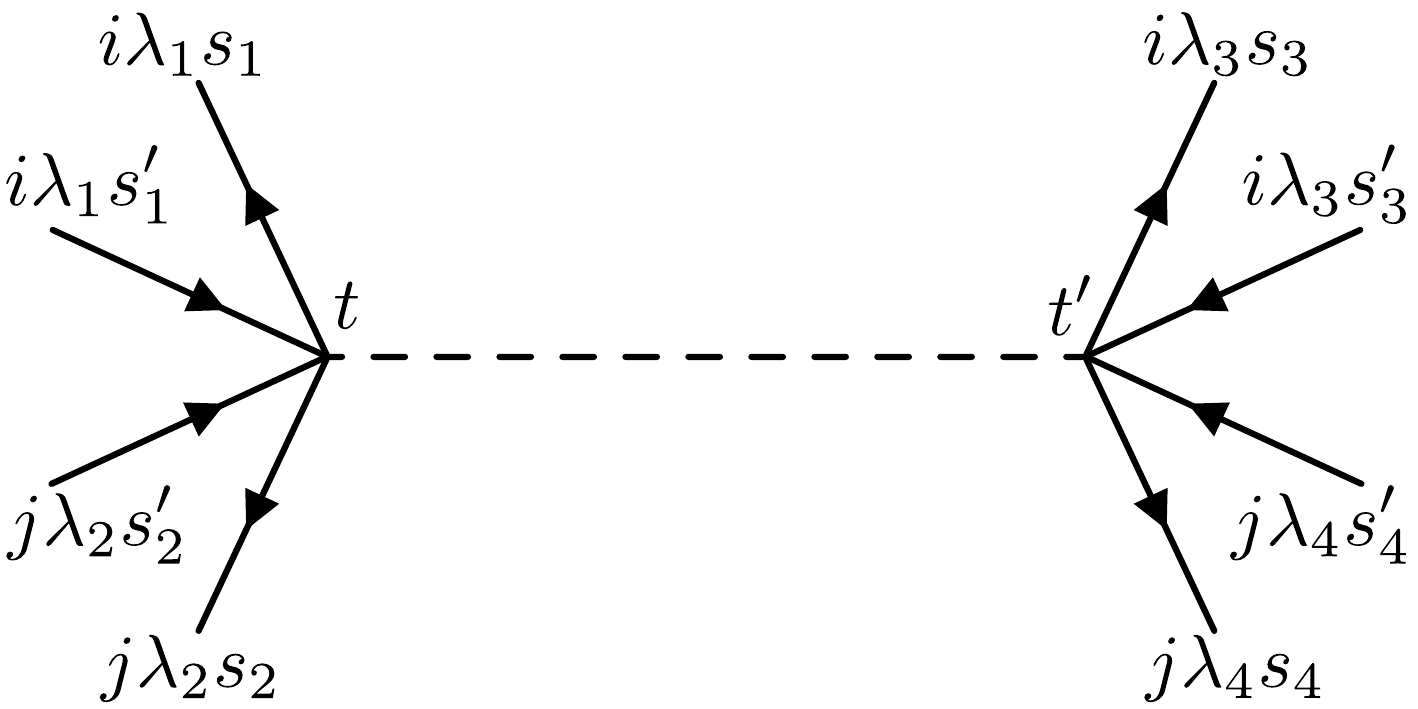}
        \caption{The elementary vertex for the disorder-averaged interaction.}
        \label{fig:basic_vertex}
    \end{figure}

    We use the Keldysh path-integral representation of the problem, which takes the following form after disorder averaging
    \begin{equation}\label{eq:S_ferm}
        S = S_\mathrm{free} + S_\mathrm{int},
    \end{equation}
    where
    \begin{equation}\label{eq:S_ferm_free}
        S_\mathrm{free} = \sum_{i,s,\lambda}\oint dt \, \bar{\psi}_{i\lambda s}i\partial_t\psi_{i\lambda s},
    \end{equation}
    is the free fermion action and
    \begin{multline}\label{eq:S_ferm_int}
        S_\mathrm{int} = \frac{iJ^2}{4N\mathcal{N}_f^2}\sum_{i>j}\sum_{\alpha\beta}\sum_{\{\lambda_i\}}\sum_{\{s_i,s'_i\}}\oint \oint dt dt' \bigg[\Big(\psi^\dagger_{i\lambda_1 s_1}(t)\boldsymbol{\sigma}^\alpha_{s_1 s_1'}\psi_{i\lambda_1 s'_1}(t)\Big)\Big(\psi^\dagger_{j\lambda_2 s_2}(t)\boldsymbol{\sigma}^\alpha_{s_2 s_2'}\psi_{j\lambda_2 s'_2}(t)\Big) \bigg]\\ \times \bigg[\Big(\psi^\dagger_{i\lambda_3 s_3}(t')\boldsymbol{\sigma}^\beta_{s_3 s_3'}\psi_{i\lambda_3 s'_3}(t')\Big)\Big(\psi^\dagger_{j\lambda_4 s_4}(t')\boldsymbol{\sigma}^\beta_{s_4 s_4'}\psi_{j\lambda_4 s'_4}(t')\Big)\bigg]
    \end{multline}
    is the interaction part, shown diagrammatically in Fig.~\ref{fig:basic_vertex}.

    \section{Derivation of the Luttinger-Ward functional for the model}

    \begin{figure}
        \centering
        \includegraphics[height=0.14\linewidth]{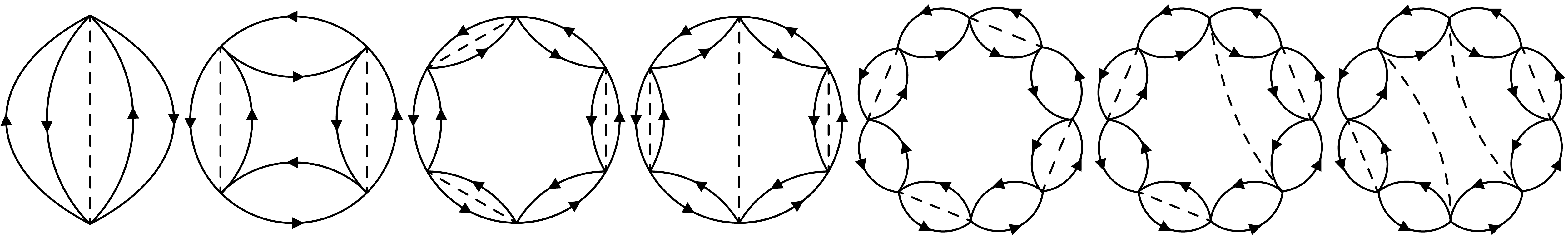}
        \caption{The first seven NLO diagrams of the LW functional.}
        \label{fig:panel_NLO}
    \end{figure}

    To proceed, we obtain the Luttinger-Ward functional, also known as the two-particle irreducible (2PI) effective action using a power expansion in $\mathcal{N}_f$~\cite{Rammer_2007}. In the first step, and before doing any approximations, we introduce bi-local source fields coupled to fermion bilinears as
    \begin{equation}
        S_\mathrm{source}=\iint dt\,dt'\, \Tr(\bm{\Psi}^\dagger(t)\cdot\bm{K}(t,t')\cdot \bm{\Psi}(t')),
    \end{equation}
    where the trace is over site, flavor and spin indices. We then define the partition functional
    \begin{equation}
        Z[\bm{K}]=\int \mathcal{D}[\bm{\Psi},\bm{\Psi}^\dagger]\,e^{i S + i S\mathrm{source}},
    \end{equation}
    and the generating functional
    \begin{equation}
        F[\bm{K}]=-i\ln{Z[\bm{K}]}.
    \end{equation}
    We define the fermionic Green's function on the Keldysh contour
    \begin{equation}
        \bm{G}(t,t')=-i\langle T_c \bm{\Psi}(t)\bm{\Psi}^\dagger(t')\rangle,
    \end{equation}
    where $T_c$ is the contour-ordering operator. One can show that
    \begin{equation}\label{eq:G_from_gener}
        \bm{G}(t,t')=-i\frac{\delta F}{\delta \bm{K}^\mathrm{T}(t',t)}.
    \end{equation}
    Eq.~\eqref{eq:G_from_gener} allows us to perform a Legendre transform
    \begin{equation}
        \Gamma[\bm{G}]=F[\bm{K}]-i\iint dt dt'\, \Tr(\bm{G}^\mathrm{T}(t,t')\cdot \bm{K}(t',t)),
    \end{equation}
    which yields
    \begin{equation}
        \frac{\delta \Gamma}{\delta \bm{G}(t,t')} = -i \bm{K}^\mathrm{T}(t',t).
    \end{equation}
    In the absence of sources $\bm{K}=0$ we get
    \begin{equation}
        \frac{\delta \Gamma}{\delta \bm{G}(t,t')}=0,
    \end{equation}
    which is the Euler-Lagrange equation for the Green's function.
    \\~
    
    The LW can always be written in the following form~\cite{Rammer_2007}:
    \begin{equation}
        \Gamma = \Gamma_\mathrm{free} + \Gamma_\mathrm{int},
    \end{equation}
    where the free part originates from the quadratic part of the action and is given by
    \begin{equation}\label{eq:Gamma_LO}
        \Gamma_\mathrm{free} = i\Tr \ln \bm{G} - i \Tr\,(\bm{G}_0^{-1}\bm{G}),
    \end{equation}
    where
    \begin{equation}
        \Big(\bm{G}_0^{-1}\Big)_{i\lambda s, j \lambda' s'}(t,t')=\delta_{ij}\delta_{\lambda \lambda'}\delta_{ss'}\delta_c(t,t') i\partial_{t'}  .
    \end{equation}
    It is clear that $\Gamma_\mathrm{free}\propto N\mathcal{N}_f$. The interacting part ($\Gamma_\mathrm{int}$) is given by the sum of all 2-particle-irreducible (2PI) diagrams made from the interaction vertex in Eq.~\eqref{eq:S_ferm_int}. We show below that it can be expanded in powers of $N$ and $\mathcal{N}_f$. Ignoring sub-extensive terms in the system size, since we work in the thermodynamics limit, we can expand $\Gamma_\mathrm{int}$ as
    \begin{equation}\label{eq:Gamma}
        \Gamma_\mathrm{int} = \underbrace{\Gamma_\mathrm{NLO}}_{O(N\mathcal{N}_f^0)} + \underbrace{\Gamma_\mathrm{NNLO}}_{O(N\mathcal{N}_f^{-1})} + O(\mathcal{N}_f^{-2}),
    \end{equation}
    Symmetries of the problem require $G$ to be diagonal in the site and flavor basis. As a result, interaction terms scale as $\Gamma_\mathrm{NLO}\sim N$, and $\Gamma_\mathrm{NNLO}\sim N\mathcal{N}_f^{-1}$.
    
 \begin{figure}
        \centering
        \includegraphics[height=0.14\linewidth]{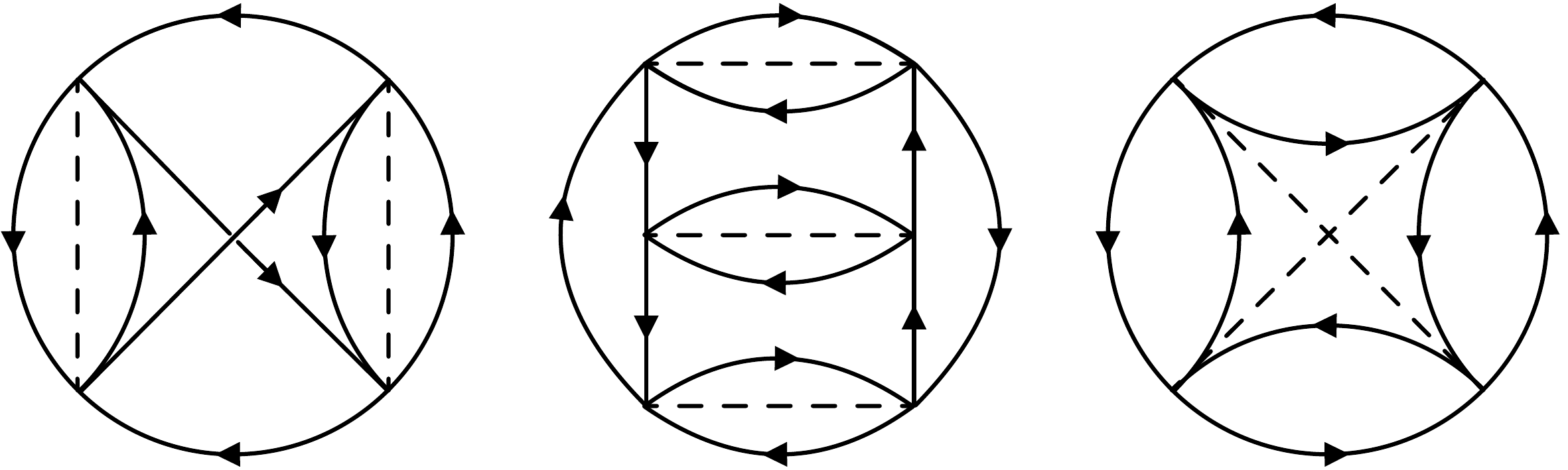}
        \caption{(Left and middle) Two of the NNLO diagrams for the LW functional. (Right) A diagram that vanishes in the thermodynamic limit.}
        \label{fig:panel_NNLO}
    \end{figure}

    It is clear from Eq.~\eqref{eq:Gamma_LO} that the leading-order approximation generates free fermion dynamics. Therefore, if we aim to capture anything non-trivial, it is necessary to include some of the higher order contributions, which can be represented in terms of 2PI Feynman diagrams. The NLO contribution is given by ring diagrams with non-crossing disorder lines, as shown in Fig.~\ref{fig:panel_NLO}. We note that there are an infinite number of NLO diagrams, at increasingly higher orders of the interaction strength, as specified by the number of disorder lines in each term. Accordingly, our treatment \emph{is non-perturbative in $J$}, a necessary condition to capture glassy physics where interactions dominate. One can attempt to sum all of the NLO diagrams as a geometric series in powers of the loops. However, such procedure is not as straightforward as in scalar field theories, due to different possible configurations of disorder lines for the same number of loops. The latter leads to different combinatorial factors that complicate the summation. Despite these difficulties, it is possible to perform a full summation of NLO diagrams using a Hubbard-Stratonovich transformation, as we show later.  We have also shown some of the NNLO and sub-extensive diagrams in Fig.~\ref{fig:panel_NNLO}. We will not take NNLO and higher order diagrams into account, as we expect them to only quantitatively modify the results.

    Having the LW functional, we can obtain the equation of motion for the renormalized Green's function using
    \begin{equation}\label{eq:ferm_dyson}
        \frac{\delta \Gamma}{\delta \bm{G}^\mathrm{T}}=0 \to \bm{G}^{-1}=\bm{G}_0^{-1}-\bm{\Sigma},
    \end{equation}
    where $\Sigma$ is the self-energy, which, at NLO, is given by
    \begin{equation}
        \bm{\Sigma} = -i \frac{\delta \Gamma_\mathrm{NLO}}{\delta \bm{G}^\mathrm{T}}.
    \end{equation}
   The self-energy diagrams corresponding the first few terms of $\Gamma_\mathrm{NLO}$ are shown in Fig.2 of the main text.

    \section{Hubbard-Stratonovich transformation}

\begin{figure}[!b]
        \centering
        \includegraphics[height=0.14\linewidth]{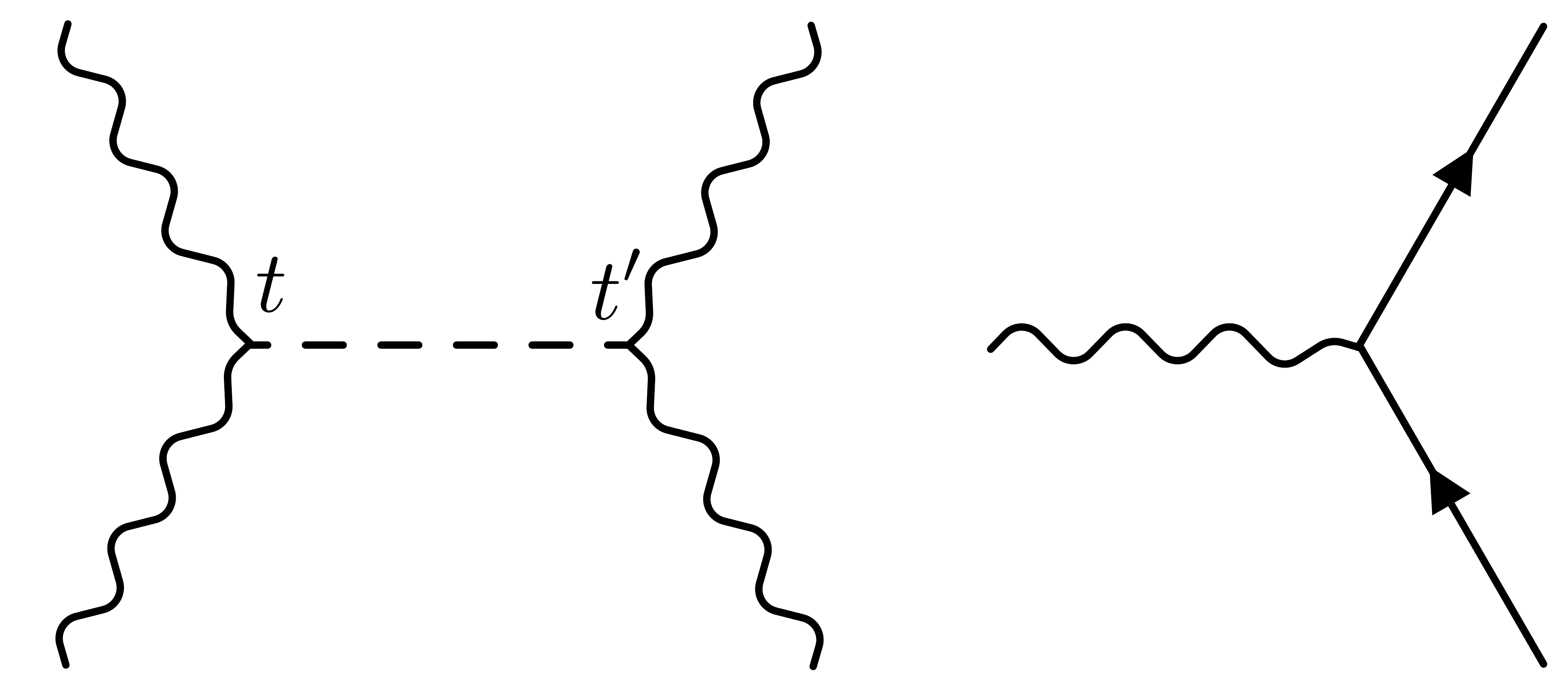}
        \caption{Two interactions vertices after the Hubbard-Stratonovich transformation.}
        \label{fig:vertices}
    \end{figure}

    The starting point is to write the action in Eq.~\eqref{eq:S_ferm} in terms of spin fields using a Lagrange-multiplier field as

\begin{equation}\label{eq:S_phi_psi}
        S = \sum_{i,s,\lambda}\oint dt \, \bar{\psi}_{i\lambda s}i\partial_t\psi_{i\lambda s}-\frac{1}{\sqrt{N}}\sum_{i>j} \oint dt\, J_{ij}  \bm{S}_i \bm{S}_j  - \sum_i \oint dt\, \boldsymbol{\Phi}_i\cdot \qty(\boldsymbol{S}_i - \frac{1}{2\sqrt{\mathcal{N}_f}}\sum_\lambda\sum_{ss'}\bar{\psi}_{i\lambda s} \boldsymbol{\sigma}_{ss'}\psi_{i\lambda s'}).
    \end{equation}
    Where $\bm{S}$ and $\bm{\Phi}$ are respectively the spin and Lagrange-multiplier fields, and the disorder average is now given by
    \begin{equation}
        \overline{J_{ij}^2}=J^2.
    \end{equation}
    Again, we can perform the disorder averaging to get the following effective interaction
    \begin{equation}\label{eq:S_J}
        S_J= \frac{iJ^2}{2N}\sum_{i>j}\oint\oint dt\,dt'\, \qty( \bm{S}_i(t)\bm{S}_j(t))\,\qty(\bm{S}_i(t')\bm{S}_j(t')).
    \end{equation}
    Due to the linear coupling between $\bm{S}$ and $\bm{\Phi}$, it is natural to express them as the components of a larger vector field
    \begin{equation}
        \bm{\Theta}_i = \begin{pmatrix}
            \bm{S}_i \\ \bm{\Phi}_i
        \end{pmatrix},
    \end{equation}
    which the following Green's function matrix
    \begin{equation}
        \bm{\Lambda}(t,t')=-i \langle \bm{\Theta}(t) \bm{\Theta}^\mathrm{T}(t')\rangle = \begin{pmatrix}
            \bm{D}(t,t') & \bm{U}(t,t') \\ \bm{V}(t,t') & \bm{W}(t,t')
        \end{pmatrix},
    \end{equation}
    where
    \begin{align}
        \boldsymbol{D}(t,t')&=-i \langle\boldsymbol{S}(t)\boldsymbol{S}^T(t')\rangle,\\
        \boldsymbol{W}(t,t')&=-i \langle\boldsymbol{\Phi}(t)\boldsymbol{\Phi}^T(t')\rangle,\\
        \boldsymbol{U}(t,t')&=-i \langle\boldsymbol{S}(t)\boldsymbol{\Phi}^T(t')\rangle,\\
        \boldsymbol{V}(t,t')&=-i \langle\boldsymbol{\Phi}(t)\boldsymbol{S}^T(t')\rangle,
    \end{align}
    which are shown by wavy lines. The vertices corresponding to the interactions in Eqs.~\eqref{eq:S_phi_psi}~and~\eqref{eq:S_J} are shown respectively in the left and right panels of Fig.~\ref{fig:vertices}.

    \subsection{Effective action after Hubbard-Stratonovich Transformation}
\begin{figure}[!t]
        \centering
        \includegraphics[height=0.14\linewidth]{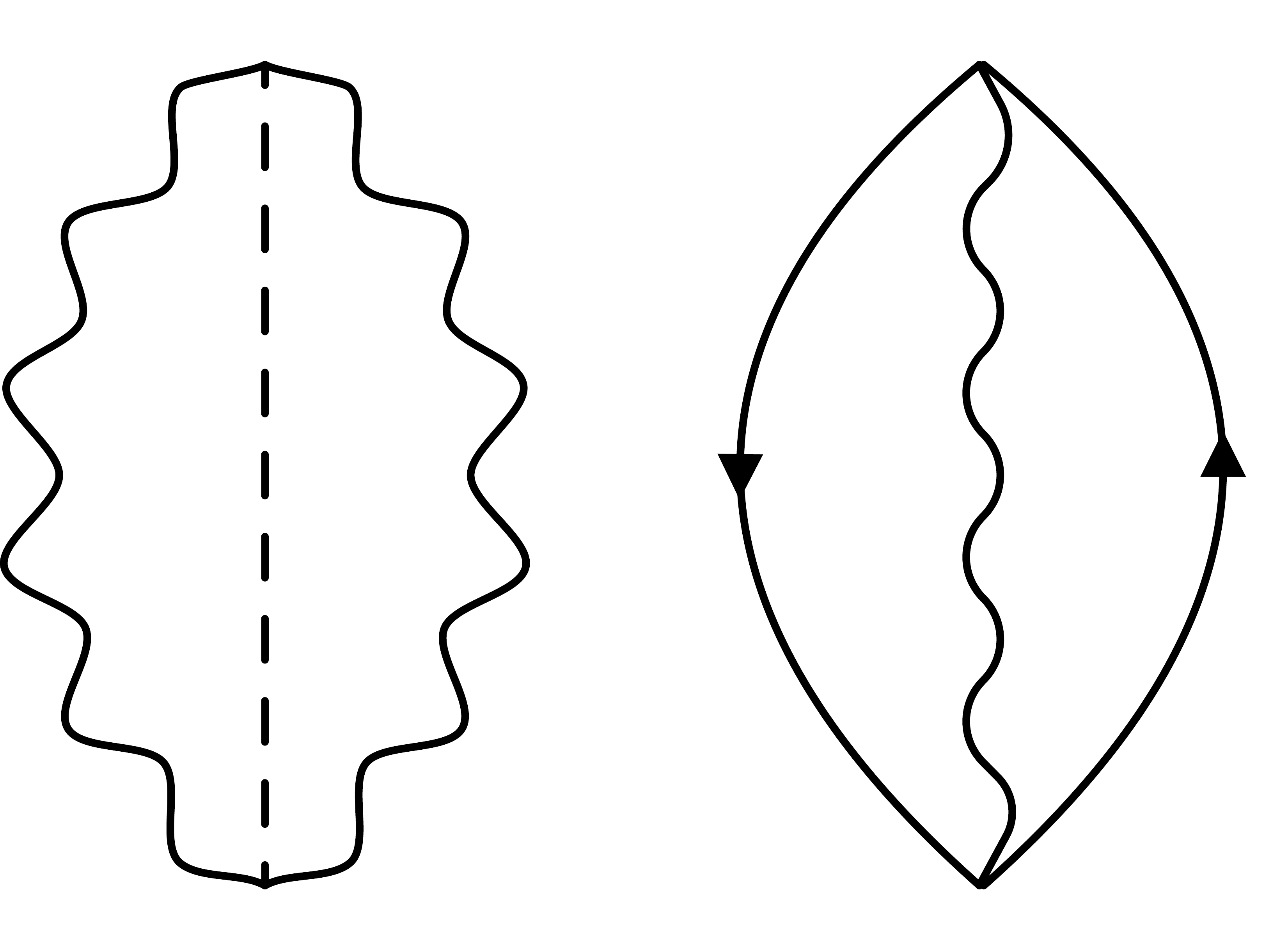}
        \caption{NLO diagrams in terms of spin Green's functions.}
        \label{fig:hs_NLO}
    \end{figure}

    We are now in the position to write down the effective action, using the same approach as before, but now we also add sources coupled to bosonic bilinears, and take the Legendre transformation with respect to both fermionic and bosonic sectors. The effective action has the same structure as in Eq.~\eqref{eq:Gamma}, with $\Gamma_\mathrm{free}$ given by Eq.\eqref{eq:Gamma_LO}. The NLO part now includes the free part for bosonic fields, which scales as $N\mathcal{N}_f^0$ since the number of boson flavors does not scale with $\mathcal{N}_f$:
    \begin{equation}
        \Gamma_\mathrm{NLO}[\bm{G},\bm{\Lambda}]=-\frac{i}{2}\Tr \ln \bm{\Lambda} + \frac{i}{2} \Tr\,(\bm{\Lambda}_0^{-1}\bm{\Lambda}) + \tilde{\Gamma}_\mathrm{NLO},
    \end{equation} where 
    \begin{equation}
        \bm{\Lambda}_0(t,t') = \begin{pmatrix}
            0 & \mathbb{I} \\ \mathbb{I} & 0
        \end{pmatrix}\delta_c(t,t').
    \end{equation}
    $\tilde{\Gamma}_\mathrm{NLO}$ contains two contributions shown in Fig.~\ref{fig:hs_NLO}, which are given by
    \begin{equation}
        \tilde{\Gamma}_{\mathrm{NLO},1} = -\frac{iJ^2}{2\mathcal{N}_f}\sum_{i>j} \sum_{\alpha \beta} \oint \oint dt\, dt' D_i^{\alpha \beta}(t,t') D_j^{\beta \alpha}(t,t'),
    \end{equation}
    \begin{equation}
        \tilde{\Gamma}_{\mathrm{NLO},2}=\frac{1}{8\mathcal{N}_f}\sum_{i}\sum_{\lambda}\sum_{\alpha\beta}\oint \oint dt\,dt' W_i^{\alpha\beta}(t,t') \Tr(\bm{\sigma}^\alpha \bm{G}_{i,\lambda}(t,t')\bm{\sigma}^\beta \bm{G}_{i,\lambda}(t',t)),
    \end{equation}
    where we have used the fact that all Green's functions are diagonal in the site basis (in the thermodynamic limit) and in the fermion flavor space.

    \subsection{Equations of motion}

    With the NLO effective action at hand, we can proceed to find the equations of motion by taking functional derivatives of $\Gamma$ with respect to Green's functions. For fermions, we get Eq.~\eqref{eq:ferm_dyson} with the self-energy given by
    \begin{equation}
        \bm{\Sigma}(t,t')=\frac{i}{4\mathcal{N}_f}\sum_{\alpha \beta}W^{\alpha\beta}(t,t') \, \bm{\sigma}^\alpha \bm{G}(t,t') \bm{\sigma}^\beta.
    \end{equation}
    For bosons, we obtain
    \begin{equation}\label{eq:boson_dyson}
        \bm{\Lambda}^{-1}-\bm{\Lambda}_0^{-1}=\begin{pmatrix}
            -\boldsymbol{\Omega}(t,t') & 0\\ 0 & -\boldsymbol{\Pi}(t,t')
        \end{pmatrix},
    \end{equation}
    where
    \begin{equation}
        \boldsymbol{\Omega}(t,t')=J^2 \boldsymbol{D}(t,t'),
    \end{equation}
    and 
    \begin{equation}\label{eq:def_Pi}
        \Pi^{\alpha \beta}(t,t')=-\frac{i}{4}\mathrm{Tr}\,\qty(\bm{\sigma}^\alpha \bm{G}(t,t')\bm{\sigma}^\beta \bm{G}(t',t)),
    \end{equation}
    is the polarization function. In the following, we safely assume SU(2) symmetry, which makes all Green's functions and self-energies diagonal in the spin basis. Therefore we get
    \begin{equation}\label{eq:SigmaR_t}
        \Sigma(t,t')=\frac{3i}{4\mathcal{N}_f}W(t,t')G(t,t'),
    \end{equation}
    \begin{equation}\label{eq:Pi_t}
        \Pi(t,t')=-\frac{i}{4}G(t,t')G(t',t),
    \end{equation}
    \begin{equation}
        \Omega(t,t')=J^2 D(t,t').
    \end{equation}

    \subsection{Equations in R/A/K Basis}
    To proceed, we expand the equations of motion in the R/A/K basis in terms of the Green's functions of classical and quantum fields~\cite{kamenev2023field}
    \begin{equation}
        \psi_{c}=\frac{1}{\sqrt{2}}(\psi_+ + \psi_-), \quad \psi_{q}=\frac{1}{\sqrt{2}}(\psi_+ - \psi_-),
    \end{equation}
    with similar definitions for $(S,\phi)$. Then, and after going to frequency domain, the retarded fermion Green's function satisfy
    \begin{equation}\label{eq:Dyson_GR}
       G^R(\omega)=\frac{1}{\omega -\Sigma^R(\omega)},
    \end{equation}
    with
    \begin{equation}\label{eq:SigR_def}
        \Sigma^R(\omega)=\frac{3i}{8\mathcal{N}_f}\int_{-\infty}^{+\infty}  \big(W^K(\nu)G^R(\omega-\nu)+ W^R(\nu)G^K(\omega-\nu)\big)\frac{d\nu}{2\pi}.
    \end{equation}
    $G^K(\omega)$ is found via the fluctuation-dissipation theorem (FDT)
    \begin{equation}\label{eq:FDT_Ferm}
        G^K(\omega)=2i\Im G^R(\omega)\tanh(\frac{\omega-\mu}{2T}),
    \end{equation}
    where at half-filling we have $\mu=0$. For retarded boson Green's functions we have
    \begin{equation}
        \begin{pmatrix} D^R(\omega) & U^R(\omega) \\ V^R(\omega) & W^R(\omega)\end{pmatrix} = -\begin{pmatrix}
\Omega^R(\omega) & -1 \\ -1  & \Pi^R(\omega)
\end{pmatrix}^{-1} ,
    \end{equation}
    which gives
    \begin{equation}\label{eq:DR_w}
        D^R(\omega) = \frac{\Pi^R(\omega)}{1-J^2 \Pi^R(\omega)D^R(\omega)},
    \end{equation}
    \begin{equation}\label{eq:WR_w}
        W^R(\omega) = \frac{J^2D^R(\omega)}{1-J^2 \Pi^R(\omega)D^R(\omega)},
    \end{equation}
    in terms of the retarded boson self-energy
    \begin{equation}\label{eq:PiR_def}
        \Pi^R(\omega)=-\frac{i}{4}\int_{-\infty}^{+\infty}  \big(G^K(\nu)G^A(\nu-\omega)+ G^R(\nu)G^K(\nu-\omega)\big)\frac{d\nu}{2\pi}.
    \end{equation}
    The Keldysh components of bosonic Green's functions can be obtained from
    \begin{equation}
        \begin{pmatrix} D^K & U^K \\ V^K & W^K\end{pmatrix} =\begin{pmatrix} D^R & U^R \\ V^R & W^R\end{pmatrix} \begin{pmatrix} \Omega^K & 0 \\ 0 & \Pi^K\end{pmatrix}  \begin{pmatrix} D^A & U^A \\ V^A & W^A\end{pmatrix},
    \end{equation}
    leading to
    \begin{align}
        D^K(\omega)&=J^2|D^R(\omega)|^2 D^K(\omega) + \frac{\Pi^K(\omega)}{|1-J^2D^R(\omega)\Pi^R(\omega)|^2},\label{eq:DK_def}\\
        W^K(\omega)&=J^4\abs{D^R(\omega)}^2\Pi^K(\omega)+\frac{J^2 D^K(\omega)}{\abs{1-J^2\Pi^R(\omega)D^R(\omega)}^2}.\label{eq:WK_def}
    \end{align}
    At thermal equilibrium, the $\Pi^K$ follows from FDT in Eq.~\eqref{eq:FDT_Ferm}
    \begin{equation}\label{eq:FDT_spin}
        \Pi^K(\omega)=2i\, \mathrm{Im}\,\Pi^R(\omega)\coth(\frac{\omega}{2T}).
    \end{equation}
    Identifying $D^R(\omega)=\chi(\omega)$ and $D^K(\omega)=-iC(\omega)$, we get the equations of motion given in the main text.

    \section{Fermion and Spin Spectral Densities above the Spin-glass transition}
    
    The imaginary part of Eq.~\eqref{eq:PiR_def}, after using Eq.~\eqref{eq:FDT_Ferm} is given by
    \begin{equation}\label{eq:PiR_Im}
        \Im \Pi^R(\omega)=-\frac12 \int_{-\infty}^{+\infty} \Im G^R(\nu)\Im G^R(\nu-\omega) \bigg[ \tanh (\frac{\nu}{2T}) - \tanh (\frac{\nu-\omega}{2T})\bigg]\,\frac{d\nu}{2\pi}.
    \end{equation}
    As long as $\Im G^R(\omega)$ is a sufficiently well-behaving function, the leading order contribution for small $\omega$ comes from the expansion of the thermal part in the square brackets
    \begin{equation}\label{eq:PiR_Im_smallw}
        \Im \Pi^R(\omega)=-\frac{\omega}{4T} \int_{-\infty}^{+\infty} \qty(\frac{\Im G^R(\nu)}{\cosh(\nu/2T)})^2\,\frac{d\nu}{2\pi}.
    \end{equation}
    In the classical paramagnetic regime, we assume that $T$ is much larger than the fermionic broadening. Therefore, we can approximately take $\cosh(\nu/2T)\approx 1$. Using the following ansatz for the fermionic Green's function
    \begin{equation}\label{eq:GR_PM}
        G^R(\omega)=\frac{1}{\omega - i\gamma},
    \end{equation}
    we obtain $ \Im \Pi^R(\omega)=-\frac{\omega}{16\gamma T}$ which is Eq.~(12) of the main text. The real part of $\Pi^R(\omega)$ can be obtained from
    \begin{equation}
        \Re \Pi^R(\omega)=\int_{-\infty}^{+\infty} \Re G^R(\nu) \Im G^R(\nu-\omega) \tanh(\frac{\nu-\omega}{2T})\,\frac{d\nu}{2\pi}.
    \end{equation}
    Using Eq.~\eqref{eq:GR_PM} and expanding to the leading order in $T^{-1}$ we get $\Re \Pi^R(0)=-1/8T$. Together, we have the following low-frequency behavior for $\Pi^R(\omega)$
    \begin{equation}\label{eq:Pi_PM}
        \Pi^R(\omega) \approx -\frac{1}{8T} \qty(1+\frac{i\omega}{2\gamma}).
    \end{equation}
    We see that $\Pi^R(\omega)$ is suppressed at high-temperatures and therefore, the factor of $J\Pi^R(\omega)$ in the denominator of Eqs.~\eqref{eq:DR_w} and~\eqref{eq:WR_w} is small  and we can approximately take $W^R(\omega)\approx J^2\Pi^R(\omega)$. In order to determine $\gamma$, we use Eq.~\eqref{eq:SigR_def} for the imaginary part of $\Sigma^R(\omega)$ for $\omega =0$
    \begin{equation}
        \Im \Sigma^R(0)=-\frac{3}{4\mathcal{N}_f}\int_{-\infty}^{+\infty}  \Im G^R(\nu) \Im \Pi^R(\nu) \bigg[ \coth(\frac{\nu}{2T}) - \tanh(\frac{\nu}{2T})\bigg]\,\frac{d\nu}{2\pi}.
    \end{equation}
    At high temperatures we can use $\tanh(\nu/2T) \ll \coth(\nu/2T)\approx 2T/\nu$. After substituting Eqs.~\eqref{eq:GR_PM} and~\eqref{eq:Pi_PM} we obtain
    \begin{equation}
        \Im \Sigma^R(0) = -\frac{3}{64\mathcal{N}_f\gamma}.
    \end{equation}
    Using $\Im \Sigma^R(0)=-\gamma$ we get $\gamma  =3^{1/2}J/8\mathcal{N}_f^{1/2}$ which is Eq.~(12) of the main text.

    \section{Connection to the SYK model}
    As was argued in the main text, away from the SG transition $J\Pi^R(\omega)$ is small and therefore, Eqs.~\eqref{eq:WR_w}~and~\eqref{eq:WK_def} can be approximated as
    \begin{equation}
        W^{R/K}(\omega) \approx J^2\Pi^{R/K}(\omega) \to W^{R/K}(t,t') \approx J^2\Pi^{R/K}(t,t').
    \end{equation}
    Upon substituting these into the fermion self-energy in Eq.~\eqref{eq:SigmaR_t} and Using Eq.~\eqref{eq:Pi_t}, we get
    \begin{equation}\label{eq:Sigma_syk}
        \Sigma(t,t')\approx\frac{3}{16\mathcal{N}_f}G(t',t)[G(t,t')]^2.
    \end{equation}
    For the complex-SYK model defined as~\cite{Chowdhury_Georges_Parcollet_Sachdev_2022}
    \begin{equation}\label{eq:H_SYK}
        H_\mathrm{SYK}=\frac{1}{(2\mathcal{N}_f)^{3/2}}\sum_{ijkl}^{\mathcal{N}_f}J_{ij;kl}\,c^\dagger_{i}c^\dagger_j c_l c_k,
    \end{equation}
    where $J_{ijkl}$ is fully antisymmetric with
    \begin{equation}
        \overline{J_{ij;kl}}=0, \qquad \overline{|J_{ij;kl}|^2}=J_\mathrm{SYK}^2.
    \end{equation}
    The self-energy of $H_\mathrm{SYK}$ in the large-$\mathcal{N}_f$ limit is given by (see Ref.\cite{Haldar_Haldar_Bera_Mandal_Banerjee_2020} for the Keldysh derivation)
    \begin{equation}
        \Sigma_\mathrm{SYK}=J^2_\mathrm{SYK} G(t',t)[G(t,t')]^2.
    \end{equation}
    Comparing this with Eq.~\eqref{eq:Sigma_syk}, we can identify $J_\mathrm{SYK}=\sqrt{3}J/4\sqrt{\mathcal{N}_f}$ which was quoted in the main text.

    \section{Spin glass phase}
    \subsection{Fermionic broadening due to spin glass order}
    The singular contribution of spin glass order to the spin correlation function given by
    \begin{equation}
        C_\mathrm{SG}(\omega)=4\pi \mathcal{N}_f q\delta(\omega),
    \end{equation}
    leads to a singular term in $W^K(\omega)$ which, according to Eq.~\eqref{eq:WK_def} is given by
    \begin{equation}
        iW_\mathrm{SG}(\omega)=16\pi \mathcal{N}_fJ^2q\delta(\omega),
    \end{equation}
    where we have used $\abs{1-J^2 \Pi^R(0)\chi(0)}=1/2$ which holds in the spin glass phase. Substituting this into Eq.~\eqref{eq:SigR_def} gives $\Sigma^R_\mathrm{SG}(\omega)=\mathcal{W}^2 G^R(\omega)$, where $\mathcal{W}=(3q)^{1/2}J$, in agreement with Eq.~(16) of the main text. For $\mathcal{N}_f\gg 1$ and $T\to 0$, we expect fluctuations to be weak and this static contribution to be dominant. Therefore, we neglect the rest of $\Sigma^R(\omega)$. This leads to the following semi-circular fermionic spectral density
    \begin{equation}
        \Im G^R(\omega)=-\frac{1}{2\mathcal{W}^2}\sqrt{4\mathcal{W}^2-\omega^2}
    \end{equation}

    \subsection{Spin spectral density in the spin glass phase}
    To find $\Im \chi(\omega)$, we need to evaluate $\Im \Pi^R(\omega)$ in the spin glass phase. We use Eq.~\eqref{eq:PiR_Im}, but take the limit $T\to 0$, which leads to
    \begin{equation}
        \Im \Pi^R(\omega) = -\frac12 \int_{0}^{\omega} \Im G^R(\nu)\Im G^R(\nu-\omega)\,\frac{d\nu}{2\pi}\approx -\frac12 \int_0^\omega \qty(\Im G^R(\nu))^2\, \frac{d\nu}{2\pi}\approx -\frac{\omega}{4\pi \mathcal{W}^2},
    \end{equation}
    giving $\Im \Pi^R(\omega)=-\alpha \omega$ where $\alpha=1/4\pi\mathcal{W}^2$. The real part of $\Pi^R(0)$ is known from the Stoner criterion 
    \begin{equation}
        J\abs{\chi(0)}=1 \to \chi(0)=-\frac{1}{J} \to \mathrm{Re}\Pi^R(0)=-\frac{1}{2J}.
    \end{equation}
    Using Eq.~\eqref{eq:DR_w} (after taking $D^R(\omega)\equiv \chi(\omega)$) we have
    \begin{equation}
        \qty(\chi(\omega)-\frac{1}{2J^2\Pi^R(\omega)})^2=\qty(2J^2\Pi^R(\omega))^{-2}-J^{-2}.
    \end{equation}
    Using $\Pi^R(\omega)\approx-1/2J-i\alpha \omega $ and $\chi(\omega)\approx -1/J + \delta \chi(\omega)$, we find
    \begin{equation}
        \qty(\delta \chi(\omega))^2\approx-\frac{4i\alpha}{J} \omega \to \Im \delta\chi(\omega) \approx -\qty(\frac{2\alpha}{J})^{1/2}\mathrm{sgn}(\omega)\sqrt{\abs{\omega}}.
    \end{equation}

    \section{Numerical solution of self-consistent equations}

    Eqs.~\eqref{eq:Dyson_GR}-\eqref{eq:FDT_Ferm},~\eqref{eq:DR_w}-\eqref{eq:PiR_def} and~\eqref{eq:DK_def}-\eqref{eq:FDT_spin} can be solved on a discrete frequency axis, by iteration starting from an initial ansatz for $G^R(\omega)$ and $\chi(\omega)$ at high temperatures. To access lower temperatures, one can use the solution for higher temperatures as the starting point of the iteration. Above the spin glass transition, all Keldysh functions, including $W^K$ and $C$,  can be obtained from the response functions using FDT as $W^K(\omega)=2i\Im W^R(\omega) \coth(\omega/2T)$ and $C^K(\omega)=-2\Im \chi(\omega) \coth(\omega/2T)$. However, in the spin glass phase, only the regular parts of $W^K$ and $C$ satisfy FDT. Therefore, one must explicitly use Eqs.~\eqref{eq:WK_def} and~\eqref{eq:DK_def} and instead use Eq.~\eqref{eq:FDT_spin}. Notably, there is no need to explicitly separate the singular part of $W^K$ and $C$ in this approach; the zero frequency components of $W^K$ and $C$ display linear scaling with the inverse of the discrete frequency spacing ($\Delta \omega$) such that
    \begin{equation}
        C(0) = \frac{4 \pi \mathcal{N}_f q}{\Delta \omega} + \tilde{C}(0),
    \end{equation}
    where we have identified $\Delta\omega^{-1} \to \delta(0)$ as the zero frequency value of the Dirac delta functions in the discrete limit. Then, the order parameter can be obtained by solving the equation for different frequency spacings and obtaining the best linear fit for $C(0)$ as a function of $\Delta \omega^{-1}$.

    \bibliography{Refs}